# Invisibility Cloaks Modeled by Anisotropic Metamaterials Based on Inductor-capacitor Networks

Xiao Liu, Chao Li*, Kan Yao, Xiankun Meng and Fang Li

*Abstract*—Based on the transformation optics, a novel transmission-line (TL) approach to realize invisibility cloaking using planar anisotropic metamaterials (MTMs) is proposed. The two-dimensional cylindrical cloaks are modeled based on inductor-capacitor (L-C) MTMs networks. The three elements of the constitutive parameters $\mu_r$, $\mu_\theta$ and $\varepsilon_z$ are all allowed to be spatially inhomogeneous which lead to the full parameter realization of a cylindrical cloak. As an example, a cloak working at VHF band is modeled and its invisibility behavior is demonstrated based on the solution of the node voltage distribution. Due to the non-resonant properties of the L-C elements, the broadband characteristic of the proposed cloaks is also evident.

## I. INTRODUCTION

Recently the use of coordinate transformation to realize electromagnetic cloaking has aroused a lot of interest. The transformation optics was proposed to control the propagation behavior of the electromagnetic wave and to achieve invisibility cloak [1], [2]. The basic principle is to manipulate the electromagnetic wave to bend around the object embedded in a cloak shell, and get back to the original propagation state after leaving the cloak shell. The possibility of the 'invisibility' has been studied intensively by analytical and numerical methods [3]-[7], such as the validity confirmation in geometric optic limit [3], the full-wave simulations with finite element and FDTD methods [4], [5], and the analytical design of cloak devices with arbitrary geometries [6], [7]. Efforts have been made to realize the cloak with complex material property, a possible option is to implement a set of reduced constitutive parameters with one component set spatially inhomogeneous while the others kept constant [8-11]. It simplifies the design but in exchange sacrifices part of the cloaking effect. The earliest realization based on split-ring resonators (SRRs) metamaterials (MTMs) in [8] is definitely a breakthrough, though due to the resonant nature of the SRRs, the bandwidth is limited. Another design in [11] offers a possible plan to realize the inhomogeneous and anisotropic cloak with layered structure of homogeneous and isotropic materials. As we know, transmission lines (TLs) are widely used because they have some great advantages, like low loss, light in weight, easy to design and most important, moderate dispersion, hence a relative large bandwidth. The TLs have been applied to achieve cloaking under a different concept which does not employ the transformation method [12], [13]. The drawback is that a large bulky object cannot be cloaked due to the periodicity of the TL network.

In this article, we present a novel two-dimensional (2D) TL approach using planar anisotropic MTMs to realize cloaking based on transformation optics. The inductor-capacitor (L-C) MTMs networks enable all three elements of the constitutive parameters ($\mu_r$, $\mu_\theta$ and $\varepsilon_z$) to be spatially inhomogeneous, so that a full parameter realization of the cylindrical cloak can be achieved. The key point of the approach is to design a TL version of anisotropic MTM which controls the three property elements independently to fulfill the specific constitutive tensors at any position in the cloak region. An implementation is carried out and the simulation results show that the wave goes around the object region with little scattering and when it returns to the background medium, the wave front can be well reconstructed over a large bandwidth.

## II. DESIGN OF THE CLOAK

### A. TL MTMs realization of the constitutive parameters

From the EM theory and the TL theory, media with specific permittivity and permeability can be modeled by L-C loaded networks. The TL versions of isotropic right-handed material (RHM) and left-handed material (LHM) have been investigated by Eleftheriades et al [14] and Caloz et al [15]. Such planar systems have been proposed to experimentally demonstrate the point-to-point focusing from a homogeneous dielectric to a LH L-C loaded TL MTM, which overcomes the classical diffraction limit [16], [17]. Additionally, Balmain et al. utilized the loaded 2D TL approach to synthesize planar anisotropic MTMs that exhibit sharp beams called "resonance cones"—a phenomenon that takes place in anisotropic plasmas that are highly resonant [18]. The dispersion relations of different kinds of planar MTMs based on L-C networks have been investigated in detail in [19].

To our best knowledge, all the previously reported anisotropic TL MTMs have permittivity and permeability tensors diagonalized in the Cartesian basis, with unit cells

Manuscript received July 27, 2009. This work was supported by the National Natural Science Foundation of China (60501018, 60990323, and 60990320), and the Knowledge Innovation Program of Chinese Academy of Sciences.

The authors are with the Institute of Electronics, Chinese Academy of Sciences, Beijing, 100190, China.
*Author to whom correspondence should be addressed. (cli@mail.ie.ac.cn)

repeated in cubic lattice. However, the medium parameters of an invisibility cloak are inhomogeneous and anisotropic with principle axes varying point to point. A correct procedure to design TL MTMs with such complex parameters requires a reasonable mapping of voltages and currents $V$ and $I$, related by the series impedance and shunt inductance, to the field quantities $E$ and $H$, related by the material parameters of the effective medium. In this paper, we consider the case of cylindrical cloaks with permittivity and permeability tensors simultaneously diagonalized in a cylindrical basis.

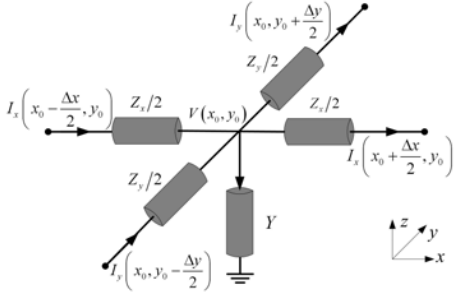

Fig. 1. Circuit model of a small section of a 2D TL

Assuming a general case of a 2D network with the unit cell schematically shown in Fig. 1, we can get following relations by applying Kirchhoff's voltage and current laws to the adjacent unit cells

$$V(x_0, y_0) - V(x_0 + \Delta x, y_0) = Z_x I_x(x_0 + \frac{\Delta x}{2}, y_0) \quad (1a)$$

$$V(x_0, y_0) - V(x_0, y_0 + \Delta y) = Z_y I_y(x_0, y_0 + \frac{\Delta y}{2}) \quad (1b)$$

$$[I_x(x_0 - \frac{\Delta x}{2}, y_0) - I_x(x_0 + \frac{\Delta x}{2}, y_0)] + [I_y(x_0, y_0 - \frac{\Delta y}{2}) - I_y(x_0, y_0 + \frac{\Delta y}{2})]$$
$$= YV(x_0, y_0) \quad (1c)$$

where $\Delta x$ and $\Delta y$ describe the dimension of the unit cell along $x$ and $y$ direction, respectively.

Also, we can write the Maxwell's equations with differential form in cylindrical coordinates as following

$$\frac{E_z(r+\Delta r, \theta) - E_z(r, \theta)}{\Delta r} = j\omega\mu_\theta H_\theta(r + \frac{\Delta r}{2}, \theta) \quad (2a)$$

$$\frac{E_z(r, \theta+\Delta\theta) - E_z(r, \theta)}{r\Delta\theta} = -j\omega\mu_r H_r(r, \theta + \frac{\Delta\theta}{2})] \quad (2b)$$

$$\frac{(r+\frac{\Delta r}{2})H_\theta(r+\frac{\Delta r}{2}, \theta) - (r-\frac{\Delta r}{2})H_\theta(r-\frac{\Delta r}{2}, \theta)}{\Delta r} -$$
$$\frac{H_r(r, \theta+\frac{\Delta\theta}{2}) - H_r(r, \theta-\frac{\Delta\theta}{2})}{\Delta\theta} = j\omega r\varepsilon_z E_z(r, \theta) \quad (2c)$$

We presume that the electric field is polarized along $z$ direction. Hence, only $\mu_r, \mu_\theta$ and $\varepsilon_z$ are relevant to the Maxwell's equations. Equations. (1) and (2) can be mapped to each other if the following replacements between the circuit and the field quantities are introduced

$$\Delta y \leftrightarrow r\Delta\theta, \Delta x \leftrightarrow \Delta r, I_y \leftrightarrow \Delta r H_r, I_x \leftrightarrow -r\Delta\theta H_\theta, V \leftrightarrow E_z d \quad (3)$$

where $d$ represents the cloak thickness along z direction. Then, the natures of the impedances and the admittances in Fig. 1 are also evident with expressions

$$Z_y = j\omega\mu_r d \frac{r\Delta\theta}{\Delta r}, \quad Z_x = j\omega\mu_\theta d \frac{\Delta r}{r\Delta\theta}, \quad Y = \frac{j\omega r\Delta\theta\Delta r\varepsilon_z}{d} \quad (4)$$

This means an inhomogeneous and anisotropic medium with permittivity and permeability diagonalized in cylindrical basis can be artificially synthesized by non-periodic TL networks. The long-wavelength approximation has to be satisfied, where the dimension of the unit cell is much smaller than the wavelength in the medium.

*B. Implementation of the cloak with L-C MTMs*

The cloak design using anisotropic L-C MTMs networks is illustrated in Fig. 2, where a cloak shell with inner radius, $a$, and outer radius, $b$, is placed in an isotropic and homogeneous background medium. Here we consider cloaking in free space.

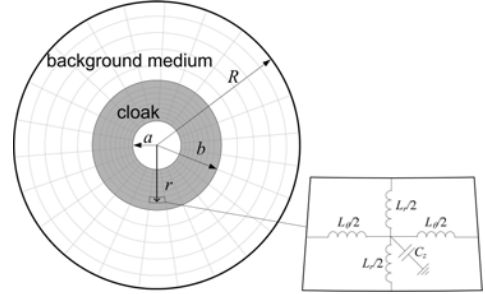

Fig. 2. Sketch map of a 2D cylindrical cloak modeled by an L-C MTM network

The relative constitutive parameters for the cloak material have been specified with the following form [8].

$$\mu_r = \frac{r-a}{r}, \quad \mu_\theta = \frac{r}{r-a}, \quad \varepsilon_z = \left(\frac{b}{b-a}\right)^2 \cdot \frac{r-a}{r} \quad (5)$$

As shown in Fig. 2, the cloak shell is then divided into small unit cells by a set of concentric circles and radial lines. Each unit cell is treated as a uniform but still anisotropic medium, whose property is represented by its geometric center. We use planar anisotropic TL structure in Fig. 1 to implement each cell, which has three branches $L_\theta, L_r,$ and $C_z,$ to control the material parameters $\mu_r, \mu_\theta$ and $\varepsilon_z,$ respectively (as shown in Fig. 2). Then the relation between the effective material parameters in (5) and the inductor and capacitor values can be derived as

$$\frac{L_\theta}{d} = \mu_r\mu_0 \frac{r\Delta\theta}{\Delta r}, \quad \frac{L_r}{d} = \mu_\theta\mu_0 \cdot \frac{\Delta r}{r\Delta\theta}, \quad C_z d = \varepsilon_z\varepsilon_0 \cdot r\Delta r\Delta\theta \quad (6)$$

where $\varepsilon_0, \mu_0$ are the permittivity and permeability of free space. As long as the dimension of each cell is small enough with respect to the wavelength, the discrete L-C model will be a good approximation to the inhomogeneous and anisotropic cloak.

We will investigate the behavior of the proposed cloak through field distributions. For the sake of simulating convenience, background medium will also be mimicked by the L-C networks aligned in the cylindrical basis. Therefore all the field information, in and outside the cloak region, can be obtained through one circuit simulation. Although in reality, the L-C cloak should be connected to the free space in a way like being embedded in a parallel plate waveguide, our circuit modeling is presented here to verify the effectiveness of the proposed approach.

The outer boundary of the free space is truncated by a circle with radius *R*, and terminated with the Bloch impedances of the adjacent unit cells to realize match absorption and to mimic the infinite extended background. Although perfect match cannot be realized through such method, the reflection at the outer boundary can be pushed small enough. Now that the constitutive parameters of the whole calculation domain are radially symmetric, only a set of L-C TL structures need to be specified along the radial direction, then we extend them to the whole domain along angular directions to complete the scheme.

## III. INVISIBILITY PROPERTIES OF THE CLOAK

Now, we consider the following design to explore the feasibility of a cloak through an L-C network. The operating frequency is chosen as 50MHz. The inner and outer radius of the cloak shell are chosen to be $a=0.5\lambda$ and $b=1.5\lambda$. The outer boundary of the background medium is chosen as $R=3\lambda$. Here $\lambda$ refers to the wavelength of 50MHz in the background medium. 15 concentric layers of L-C MTMs are used to describe the cloak shell and another 18 to describe the outside free space background. Each layer is equally divided into 90 unit cells along angular direction. The inductors $L_r$ at the inner boundary of the cloak are grounded to create an effect of a PEC boundary there, whereas those at the outer boundary of the whole networks are terminated with Bloch impedances calculated following the procedure in [20]. Based on (4) and (5), the constitutive parameters along radial directions are calculated and the parameters of the corresponding L-C MTMs networks are derived for both the cloak shell and the background medium, the results are shown in Fig. 3. In our circuit implementation, because the free space is also mimicked by an L-C network, different choices of *d* result in the proportional impedance changes for all the branches (which can be obviously concluded from (4)), hence will not interfere with the whole distribution pattern and affect the cloaking effect. In our examples, *d* is chosen as 0.6 m , which is 1/10 wavelength at 50MHz. To investigate the invisibility properties, idealized inductors and capacitors with calculated values are directly connected to each other to form the network, then the node voltages of each unit cell are solved using Agilent's Advanced Design System (ADS) circuit simulator. Fig. 4 shows the resulted voltage distributions. In Fig. 4 (a), a bare conducting cylinder with a radius *a* is placed in free space, and a point source is located in the background medium. The scattering is obvious from the simulated voltage distribution. Fig. 4 (b) shows the behavior of the same conducting cylinder surrounded by our L-C cloak. We can see that the wave travels as a normal cylindrical wave before it encounters the cloak, then it interacts with the cloak shell and goes around the hidden object with little back scattering and forward scattering. It should be mentioned that, for different kinds of terminations at the inner boundary of the cloak, the voltage distributions are almost identical which means any inclusion inside the cloak could not interfere with the total electromagnetic field outside, therefore can be successfully hidden.

Due to the non-resonant nature of TL structures, our L-C MTMs cloak may exhibit moderate dispersion, which means a comparatively broad bandwidth is expected for the proposed cloak. To verify that, we investigate the scattering power of the uncloaked conducting cylinder, $P_{unc}$, and the cloaked one, $P_c$, when illuminated by a cylindrical wave excited by a voltage source over a frequency range from 30MHz to 60MHz, with 0.5MHz interval. To calculate the scattering power, firstly the incident field is subtracted from the total field to obtain the scattering field, and then the scattering power $P$ is calculated based on

$$P = \sum_{i=1}^{90} V(i) \cdot I_{out}^{*}(i) \qquad (7)$$

where $V(i)$ and $I_{out}(i)$, $i=1,2,\cdots,90$, are the voltages and outward currents on each node along a closed contour in the background medium, respectively. The scattering power ratio of a cloaked conducting cylinder to a bare cylinder is given as a function of frequency in Fig. 5. Obviously, the L-C cloak can reduce the scattering of an object effectively over this frequency range, and if we take the ratio lower than 0.1 to mark the bandwidth $BW$, it exhibit a $BW$ of about 28.5MHz lying between 30MHz to 58.5MHz. In fact, the broadband effect of the cloak can be well extended to a much lower frequency. However we should be more careful dealing with higher frequencies, because the wavelengths become too short compared to the period of the L-C network.

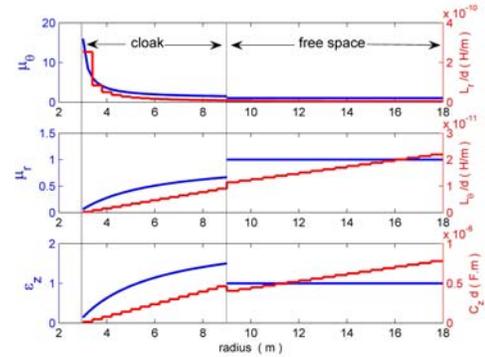

Fig. 3. The calculated effective constitutive parameters for both the cloak and free space, and the L-C MTMs parameters used to implement them.

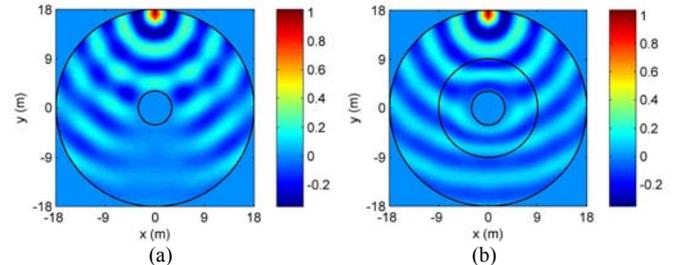

Fig. 4. Voltage distribution snapshots from ADS circuit simulations. The simulation results of (a) a bare conducting cylinder and (b) a cloaked conducting cylinder excited by a cylindrical wave with the point source located in free space. (unit: V)

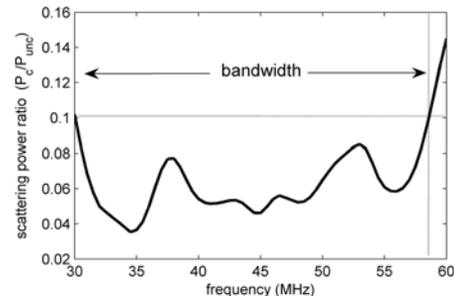

Fig. 5. The scattering power ratio of a cloaked conducting cylinder, $P_c$, and an uncloaked bare conducting cylinder, $P_{unc}$.

It should also be noted that, the singularities on the inner boundary of the cloak brought by coordinate transformation cannot be avoided, so for the TL cells involving the inner boundary, finite values of the parameters are used as approximations. Therefore, even though theoretically the field solutions are more accurate if the grids are set denser, the full-parameter cloak cannot be called an 'ideal' one. In spite of that, the TL cloak has been demonstrated to be able to greatly reduce the scattering of an object over a relatively wide frequency range. That is a great improvement on the bandwidth with respect to the cloak realization using SRRs. Besides, since the cloak structure is composed of only L-C elements, it can be easily carried out with lumped elements, which is of great use in applications.

## IV. CONCLUSION

This paper provides us with a novel TL approach to realize invisibility from incident electromagnetic wave using planar anisotropy L-C MTMs networks. The key point is to find the relationship between the parameters of the 2D anisotropy L-C MTM and the constitutive parameter tensors of a real anisotropy material. Then by arranging enough such 2D anisotropy L-C cells into an annular cloak area, with each one representing the specific property of its position, the TL cloak is realized. The simulation results demonstrate that the TL cloak can greatly reduce the scattering of an object over a large frequency range.

REFERENCES

[1] J. B. Pendry, D. Schurig, and D. R. Smith, "Controlling electromagnetic fields," *Science*, vol. 312, pp. 1780-1782, 2006.
[2] U. Leonhardt, "Optical conformal mapping," *Science*, vol. 312, pp. 1777-1780, 2006.
[3] D. Schurig, J. B. Pendry, and D. R. Smith, "Calculation of material properties and ray tracing in transformation media," *Opt. Express*, vol. 14, pp. 9794-9804, 2006.
[4] S. A. Cummer, B-I. Popa, D. Schurig, D. R. Smith, and J. B. Pendry, "Full-wave simulations of electromagnetic cloaking structures," *Phys. Rev. E*, vol. 74, p. 036621, 2006.
[5] Y. Zhao, C. Argyropoulos, and Y. Hao, "Full-wave finite-difference time-domain simulation of electromagnetic cloaking structures," *Opt. Expr.*, vol. 16, pp. 6717-6730, 2008.
[6] A. Nicolet, F. Zolla, and S. Guenneau, "Electromagnetic analysis of cylindrical cloaks of an arbitrary cross section," *Opt. Lett.*, vol. 33, pp. 1584-1586, 2008.
[7] C. Li, and F. Li, "Two-dimensional electromagnetic cloaks with arbitrary geometries," *Opt. Expr.*, vol. 16, p. 13414, 2008.
[8] D. Schurig, J. J. Mock, B. J. Justice, S. A. Cummer, J. B. Pendry, A. F. Starr, and D. R. Smith, "Metamaterial electromagnetic cloak at microwave frequencies," *Science*, vol. 314, pp. 977-980, 2006.
[9] I. I. Smolyaninov, Y. J. Hung, and C. C. Davis, "Two-dimensional metamaterial structure exhibiting reduced visibility at 500 nm," *Opt. Lett.*, vol. 33, pp. 1342-1344, 2008.
[10] W. Cai, U. K. Chettiar, A. V. Kildishev, and V. M. Shavlaev, "Optical cloaking with metamaterials," *Nature photon.*, vol. 1, pp. 224-227, 2007.
[11] Y. Huang, Y. Feng, and T. Jiang, "Electromagnetic cloaking by layered structure of homogeneous isotropic materials," *Opt. Express*, vol. 15, pp. 11133-11141, 2007.
[12] P. Alitalo, O Luukkonen, L. Jylhä, J. Venermo, and S. Tretyakov, "Transmission-line networks cloaking objects from electromagnetic fields," *IEEE Trans. Antennas Propagat.*, vol. 56, no. 2, pp. 416-424, 2008.
[13] P. Alitalo, F. Bongard, J-F. Zürcher, J. Mosig, and S. Tretyakov, "Experimental verification of broadband cloaking using a volumetric cloak composed of periodically stacked cylindrical transmission-line networks," *Appl. Phys. Lett*, vol. 94, p. 014103, 2009.
[14] G. V. Eleftheriades, A. K. Iyer, and P. C. Kremer, "Planar negative refractive index media using periodically L-C loaded transmission lines," *IEEE Trans. Microw. Theory and Tech.*, vol. 50, no. 12, pp. 2702-2712, 2002.
[15] C. Caloz, H. Okabe, T. Iwai, and T. Itoh, "Transmission line approach of left-handed (LH) materials," in *Proc.USNC/URSI National Radio Science Meeting*, SanAntonio, TX, Jun. 2002, vol. 1, p. 39.
[16] A. Grbic, and G.V. Eleftheriades, "Negative Refraction, growing evanescent waves and sub-diffraction imaging in loaded transmission-line metamaterials," *IEEE MTT*, vol. 51, no. 12, pp. 2279-2305, 2003.
[17] A. Grbic, and G. V. Eleftheriades, "Overcoming the diffraction limit with a planar left-handed transmission-line lens," *Phys. Rev. lett.*, vol. 92, no. 11, pp. 117403, 2004.
[18] K. G. Balmain, A.A.E. Luttgen, P.C. Kremer, "Power flow for resonance cone phenomena in planar anisotropic metamaterials," *IEEE Trans. Antennas Propagat.*, vol. 51, no. 10, pp. 2612-2618, 2003.
[19] Y. Feng, X. Teng, Y. Chen, and T. Jiang, "Electromagnetic wave propagation in anisotropic metamaterials created by a set of periodic inductor-capacitor circuit networks," *Phys. Rev. B*, vol. 72, p. 245107, 2005.
[20] A. Grbic and G. V. Eleftheriades, "Periodic analysis of a 2-D negative refractive index transmission line structure," *IEEE Trans. Antennas and Propagat*, vol. 51, no. 10, pp. 2604-2611, 2003.